# Five Generic Processes for Behavior Description in Software Engineering

Sabah Al-Fedaghi
Computer Engineering Department
Kuwait University
Kuwait
sabah.alfedaghi@ku.edu.kw

*Abstract*—Behavior modeling and software architecture specification are attracting more attention in software engineering. Describing both of them in integrated models yields numerous advantages for coping with complexity since the models are platform independent. They can be decomposed to be developed independently by experts of the respective fields, and they are highly reusable and may be subjected to formal analysis. Typically, behavior is defined as the occurrence of an action, a pattern over time, or any change in or movement of an object. In systems studies, there are many different approaches to modeling behavior, such as grounding behavior simultaneously on state transitions, natural language, and flowcharts. These different descriptions make it difficult to compare objects with each other for consistency. This paper attempts to propose some conceptual preliminaries to a definition of behavior in software engineering. The main objective is to clarify the research area concerned with system behavior aspects and to create a common platform for future research. Five generic elementary processes (creating, processing, releasing, receiving, and transferring) are used to form a unifying higher-order process called a thinging machine (TM) that is utilized as a template in modeling behavior of systems. Additionally, a TM includes memory and triggering relations among stages of processes (machines). A TM is applied to many examples from the literature to examine their behavioristic aspects. The results show that a TM is a valuable tool for analyzing and modeling behavior in a system.

*Keywords-conceptual modeling; process modeling; behavior; behavior modeling; elementary generic process*

## I. Introduction

Behavior modeling for system and software architecture specification is attracting more attention in software engineering [1]. For example, according to Ringert et al. [2], describing both a system's architecture and behavior in integrated models yields many advantages to coping with complexity: the models are platform independent, can be decomposed to be developed independently by experts of the respective fields, are highly reusable, and may be subjected to formal analysis. Behavior modeling involves representing different types of behavior, including internal system behavior, interaction with the environment, and collaboration between systems. There are many different approaches to modeling behavior, such as grounding it on state-transition systems and diagrammatic languages, including UML [1].

Definitions of *behavior* (of a system) are plentiful in the scientific and philosophical literature [3]. In general, the classical description of behavior [3] can be summarized as follows:

(a) Behavior is the occurrence of an action or reaction.
(b) Behavior as any change or movement of an object.

System behavior has been described using many terms, including the total movements of an agent [4], any activity [5], and a process of an inner entity causing a movement or environmental outcome [6]. In system studies, behavior refers to the interaction between the environment in which the system is embedded and the action/reaction of the system itself.

A person's behavior is typically described in terms of his or her actions. The La Mettrie doctrine [7] maintains that human behavior emerges from machines, that all motions are mechanical, and that, in this sense man, is a machine [7].

> Think of how you would describe the operation of, say, an automobile; that is, think of how the automobile operates. Excluding interactions with co-systems (like users), operation becomes a description of the internal working of the product. *Behaviour* maps inputs to outputs only. Operation describes how the inputs are turned into outputs [8]. (Italics added.)

The behavior of a system is realized by means of the behavior of its subsystems, including their interactions with each other and with the environment [9]. Accordingly, there is an internal behavior of a subsystem and the *emerged* behavior of the whole system.

Ants are relatively simple components in the complex system of the ant colony. More specifically, each ant or component's behavior is relatively simple compared to what the overall system is doing. An ant colony as a whole is capable of engaging in complex behaviors, such as building nests, foraging for food, raising aphid "livestock," waging war with other colonies, and burying their dead [10].

Often the specification of a software system's behavior is given simultaneously in different ways, such as natural language, flowcharts and pseudocodes, UML, data flow diagrams, and so forth that are encoded in a variety of programming languages [11]. One problem, in this context, is that these various descriptions make it difficult to compare them with one another for consistency. To achieve consistency, it would be useful to write one description to serve both requirements and design [11].

All behaviors involve change in time. It is said that "behavior is a function of time and structure is a function of space. [They] are intimately interlinked . . . A behavioral description presumed a structural description, but a structural description also presumed a behavioral description" [12].

In software engineering, it is typical to describe behavior as an application of the classical input–process–output model. According to Salustri [8], "The degree and extent of the response of an object's structure to a stimulus is called the object's behavior . . . An object's behavior is like 'consumption' of inputs (stimuli) and 'production' of outputs (responses). Behavior is 'what the artefact does' and it can be derived from structure" [13].

In UML, an object's behavior is defined in terms of the input/output messages. It involves the creation of multiple views, which is unavoidable [14]. A problem related to behavior modeling is that "these views, which represent different aspects of system structure and behavior, overlap, raising consistency and integration problems. Moreover, the object-oriented nature of UML set the ground for several behavioral views in UML, each of which is a different alternative for representing behavior" [14]. According to Brush [12], UML "is incapable to connect structure and behavior."

The abovementioned problems in behavior modeling point to the need for further research in this area. This paper attempts to propose some conceptual preliminaries to a definition of behavior in software engineering. The main objective is to clarify the research area concerned with system behavior aspects and to create a common platform for future research. Specifically, the paper describes a unified specification for behavior that is intended to serve both the requirements and design of software systems. We approach the topic at the conceptual level using a diagrammatic language called thinging machine (TM). The following examples from the literature gives an idea of the level of behavior modeling that is studied in this paper.

**Example**: Figs. 1 and 2 exemplify a typical method of describing the behavior of a system in terms of an algorithm, by which the behavior of a system can be modeled as a set of events [15]. Clearly, the purpose of showing these diagrams is not to present a fair discussion of their contents; rather, the goal is showing a view of the level of the involved description.

The behavior of the user, system, and environment is specified as follows:
(1) The user inputs data or submits a query and receives processed data or an error message.
(2) The application receives account data, checks for data consistency, stores account data, and sends processed data [15].

In the next section, we introduce our diagrammatic modeling language, TM, which will be used throughout the paper to analyze the notion of behavior. The TM model has been used in many papers; the references include TM-related papers published in 2018-2019 [16-40].

In a TM, we claim that the generic elementary changes or processes are of five types (called stages): creating, processing, releasing, receiving, and transferring things. These elementary processes form a complex process (abstract machine) called a TM, as shown in Fig. 3. In a TM, we assume that there are no "disembodied" elementary processes (i.e., all elementary processes are inside machines). A TM can be put into the form of the very well-known input–process–output model (Fig. 4).

A TM forms the patterns or templates of elementary processes that composes the system. They are the fundamental processes constituting a mosaic or network of machines. Additionally, the TM model includes memory and triggering (represented as dashed arrows, as will be discussed later) relations among the processes' stages (machines).

Note that a TM manifests structure and behavior simultaneously. A UML class can be represented as a TM with the attributes flowing into it. Only five elementary processes are used because they represent generic ideas, the way the three forms of water (liquid, vapor, and solid) represent three generic concepts. These elementary processes have been called different names.

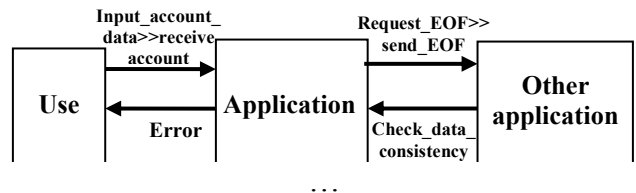

Fig. 1. Illustration of the level of the behavior description of interest in this paper (partially adapted from [15]).

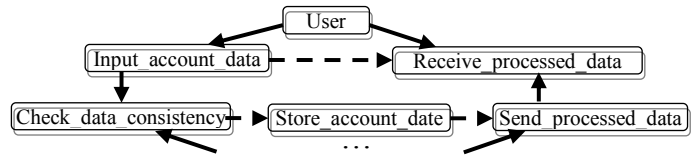

Fig. 2. Illustration of a view of an event trace (partially adapted from [15]).

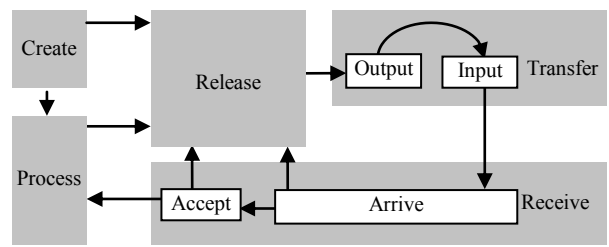

Fig. 3. Thinging machine.

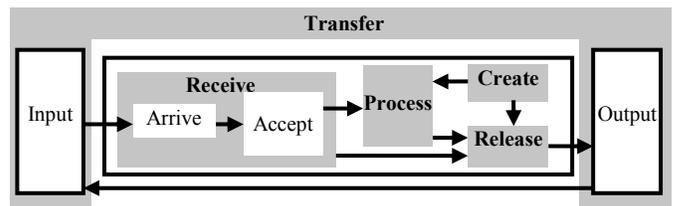

Fig. 4. Another form of description of a TM.

- *Create*: generate, produce, manufacture, give birth, initiate . . .
- *Process*: change, modify, adjust, amend, alter, revise . . .
- *Receive*: obtain, accept, collect, take, get, be given . . .
- *Release*: allow, relieve, discharge, free, acquit, clear . . .
- *Transfer*: transport, transmit, carry, communicate . . .

**Example**: Fig. 5 shows a static description of a mousetrap using a TM. The model includes a trap, a mouse, bait, and a door (circles 1, 2, 3, and 4). The door (4) is initially open (5). The bait creates a smell (6) that flows to the mouse (7) to be processed, and it triggers (8) the mouse (9) to move inside the trap (10). When the mouse is inside the trap, the door is closed (11).

Later in this paper, we will deal with the dynamic behavior of the TM model. Fig. 6 illustrates this dynamic description as a sequence of events over time: a→b→c→d. That is, (a) there is a trap, a mouse, bait, and an open door; (b) the bait creates a smell that flows to the mouse; (c) the smell triggers the mouse to enter the trap; and (d) the door is closed.

## II. TM SUBSTANTIATION

Currently, no formal proof exists that the five TM stages are sufficient to describe *all* behavior processes. The TM model has been applied to many real systems, such as in phone communication [29], physical security [17], vehicle tracking [19], intelligent monitoring [32], asset management [20], information leakage [22], engineering plants [23], inventory management processes [24], procurement processes [26], public key infrastructure network architecture [26], bank check processing [35], waste water treatment [39], computer attacks [26], provenance [42], services in banking industry architecture network [31], and digital circuits [37]. Additionally, we will expend extensive effort to partially substantiate the claim of the genericity of the TM stages in different fields.

### A. Illustration of Elementary Processes in English

According to the English language site TESOL Direct [51], *process* is a verb that indicates a change from one state to another. Verbs often signify motion, and "every motion necessarily supposes some being or existence" [52]. In addition, "they also express all the different actions and movements of all creatures and all things, whether alive or dead" [52]. In generative grammar, verbs play a central role because they function "as the nucleus in the deep structure, from which various surface utterances are processed" [52]. The concept of a verb is closely related to process; in fact, process is sometimes viewed as a type of verb, or as a series of activities (i.e., verbs). According to Cousins [53], a process is "a set of interrelated or interacting activities which transforms inputs into outputs." Usually, verbs are used to describe the steps in a process (activities), and nouns are used to describe the items output by activities to become input for other activities.

A vast amount of work has been done in the field of verb semantics. Here, we discuss only few verbs to demonstrate how a TM exemplifies them. Here are some examples from the most basic English phrases that people use every day [54] modeled in a TM.

*(a) Thanks for the birthday money.* This phrase can be modeled as shown in Fig. 7. The figure expresses that birthday money is given by you and such an act motivates me to thank you. A similar explanation can be applied to other phrases; accordingly, it is sufficient to only present some phrases and their TM diagrams.

*(b) Excuse me (to get attention).* (See Fig. 8.)

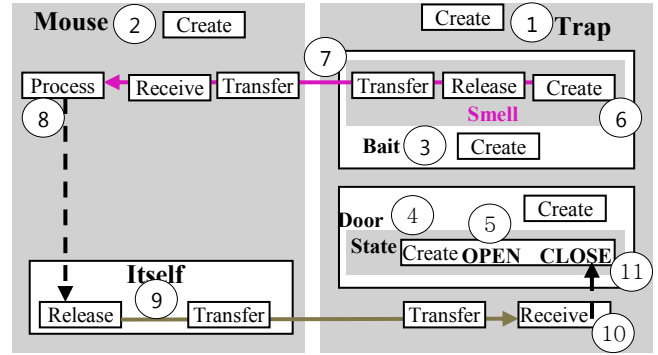

Fig. 5 A TM representation of a mousetrap.

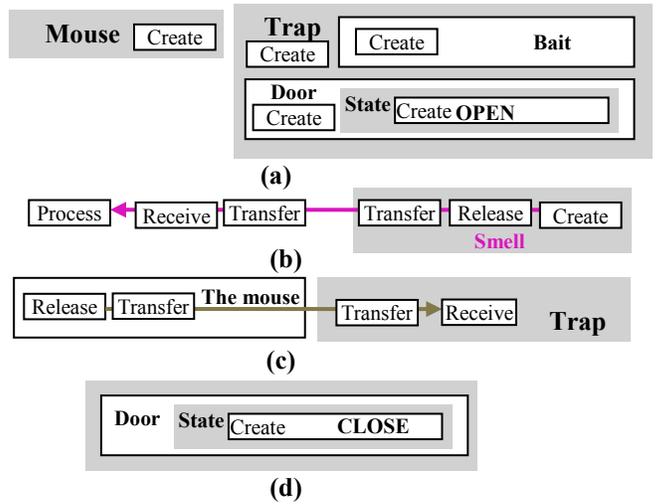

Fig. 6. Illustration of the dynamic description of the mousetrap.

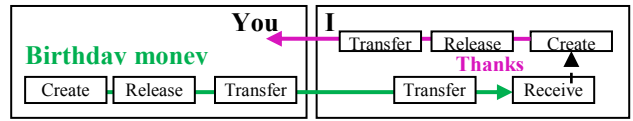

Fig. 7. The TM graph of *Thanks for the birthday money*.

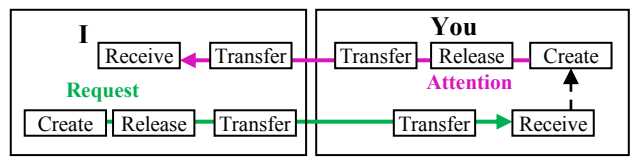

Fig. 8. The TM diagram of *Excuse me*.

*(c)* *Could you please talk slower?* (See Fig. 9.)

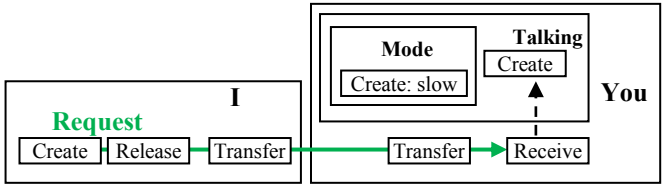

Fig. 9. The TM diagram of *Could you please talk slower?*

*(d)* *I am running a bit behind, but it will be done by noon!* (See Fig. 10.) The sentence divides the time into now, before now, and noon. Since starting the task, currently (now), its processing is late, but by noon, the processing of the task will be completed. Note that not continuing the flow of the task implies that it has finished. The figure is simplified by not including the Create stage for subject I under the assumption that the mere presence of this box implies that subject I exists. We can also put a box in I (Fig. 11) labeled "speaking" so the dark boxes in Fig. 10 become the content shown in Fig. 11. Fig. 11 shows a possible simplification of the TM diagram by deleting transfer, release, and receive under the assumption that the direction of the arrow indicates the flow.

*B. Elementary processes in process image graph*

As another example of expressing situations in terms of a TM, consider the "process image" diagram given in Sheninger [55], where the elementary TM processes express every change in the situation. Fig. 12 shows the corresponding TM representation, where every change is represented in terms of the TM stages.

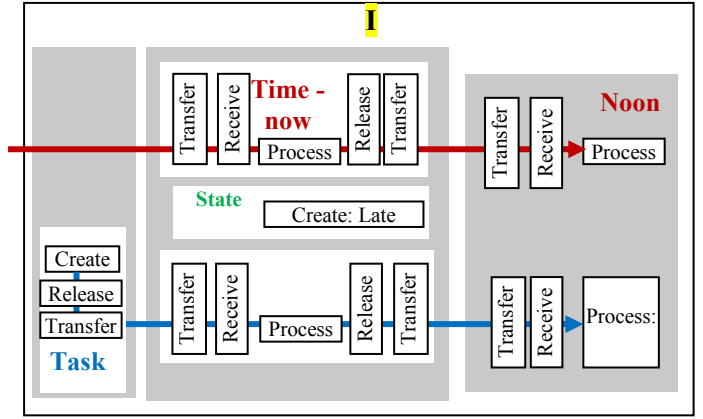

Fig. 10. The TM graph of *I am running a bit behind, but it will be done by noon!*

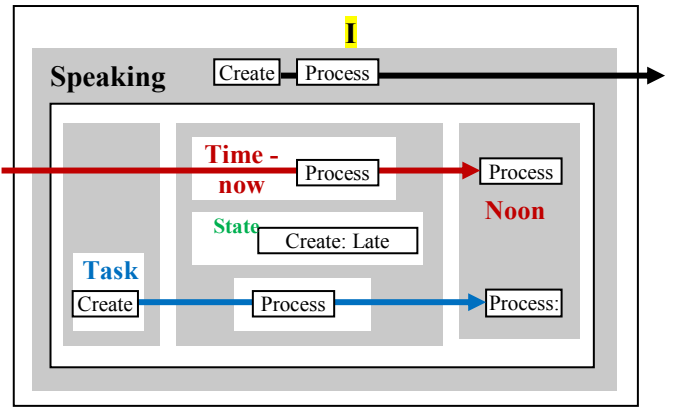

Fig. 11. Adding speaking and simplification in the TM graph of *I am running a bit behind, but they'll be done by noon!*

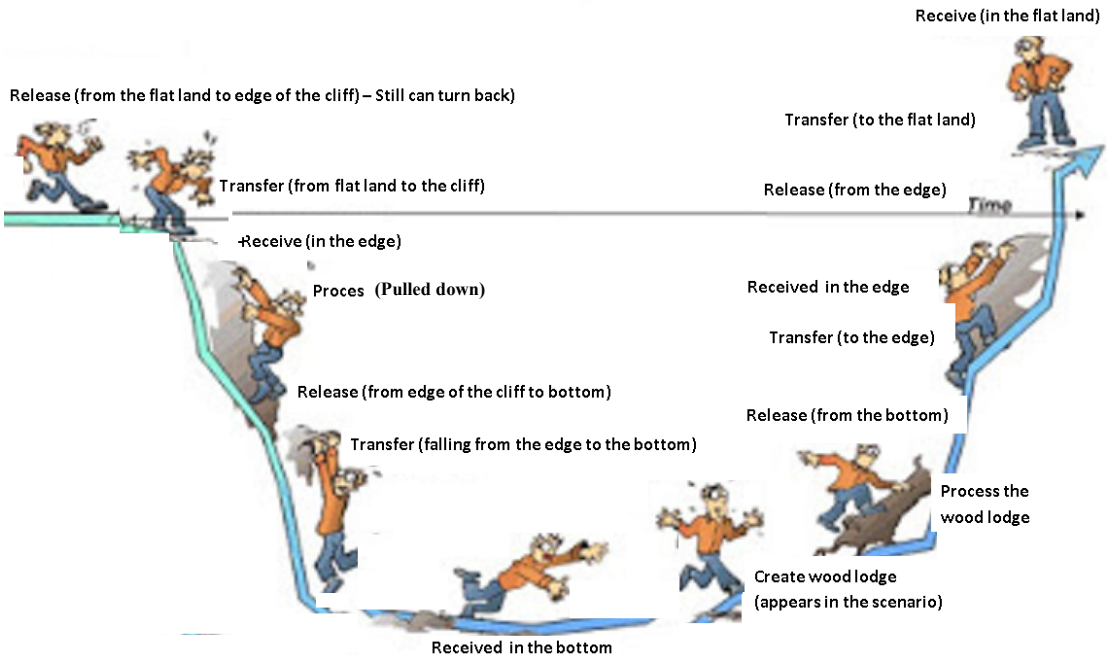

Fig. 12. Illustration of applying the five elementary processes in TM (The figure is adapted and modified from [55])

## C. Flow of Things

A thing is created, processed, released, transferred, and received. A thing is simultaneously a mechanism that creates, processes, releases, transfers, and receives; that is, a thing is a TM. No thing (nothing) means no TM (i.e., no creation, no processing, no releasing, no transferring, and no receiving).

Creation is a loaded notion that may refer to existence in our system of concern (e.g., an inventory system). For our purpose, it is sufficient to say that *creation* denotes the appearance of a thing in a system for the first time. A thing can be created by processing other things (see Fig. 13).

The flow of a thing is the change of place from one stage to another in a TM. The flow is also a mechanism of exchanging things among machines. A thing also changes when it is conceptualized as a machine and when things inside it flow through its stages (see Fig. 14).

## D. Process and Things

According to Carr [56], in discussing Bergson's ideas, movement is original, and things are derived from movement. This idea was refined by Whitehead's notion of *process* that emphasizes becoming and changing over static being. "Actuality consists not of individual objects with attributes, but rather of interwoven processes" [57]. This brings forth a related problem that illustrates the notion of things and machines (processes) in a TM.

According to Gentner [58], "As far back as Aristotle, we find arguments that the kinds of things denoted by nouns are different from, and more fundamental ontologically than, the kinds of things denoted by verbs." Gentner [48] quotes a source, "And so one might even raise the question whether the words 'to walk,' 'to be healthy,' 'to sit,' imply . . . none of them is either self-subsistent or capable of being separated from substance [which] are seen to be more real . . . for we never use the word . . . 'sitting' without implying this."

Consider the process of sitting in a TM.

Fig. 15 (a) says that John is sleeping (more familiar: John is living). The right diagram in the figure indicates the prior appearance of John (noun sense), then sleeping (verb sense) takes place.

Fig 15 (b) says that John enters (the state of) sleeping. This indicates the prior presence of sleeping

Fig. 15 (c) says that sleeping flows through John (more familiar: feeling flows through John).

Fig. 15 (d) says that sleeping embraces John.

Such alteration of occurrences of the noun and verb senses strengthens the TM's conceptualization of things and machines as two faces of the same phenomenon.

## E. TM, Change, and behaviour

This section connects change, behavior, and control in the TM model.

The TM model acts as the basin in which changes are coordinated. Transferring, releasing, receiving, processing, and creating things are changes in the TM model. In other words, the TM system (represented by the TM model) is the observer. The flow of things among the five TM stages (including triggering) makes changes. *Changes* with time make *events*. The chronology of events makes the *behavior* of the system. Control is the machine that is "looking" inward (information feedback) at the system processes and that changes their behaviors. It is an observer that collects knowledge about the behavior of other machines even though it is actually inside the system.

According to our thesis, things are processes. A thing has distinctiveness (outlines that enclose stages); thus, it can be created, released, processed, transferred, and/or received. Heidegger's thing [59] of "gathering together" its constituents, as illustrated in the bridge that makes the environment (banks, stream, and landscape) into a unified whole [59], is applied to the internal operations of the bridge as a machine that creates, processes, releases, transfers, and receives things, as illustrated in Fig. 16.

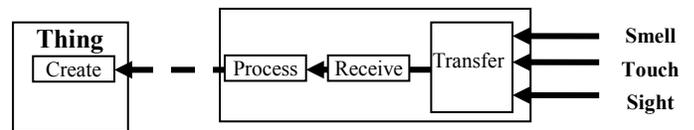

Fig. 13. A thing is created by processing other things.

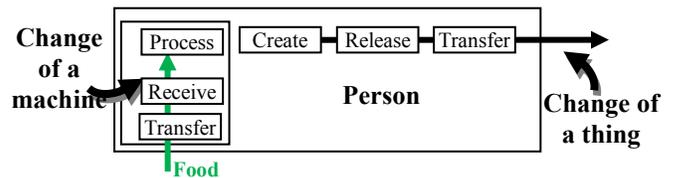

Fig. 14. Illustration of changes in machines and changes in things.

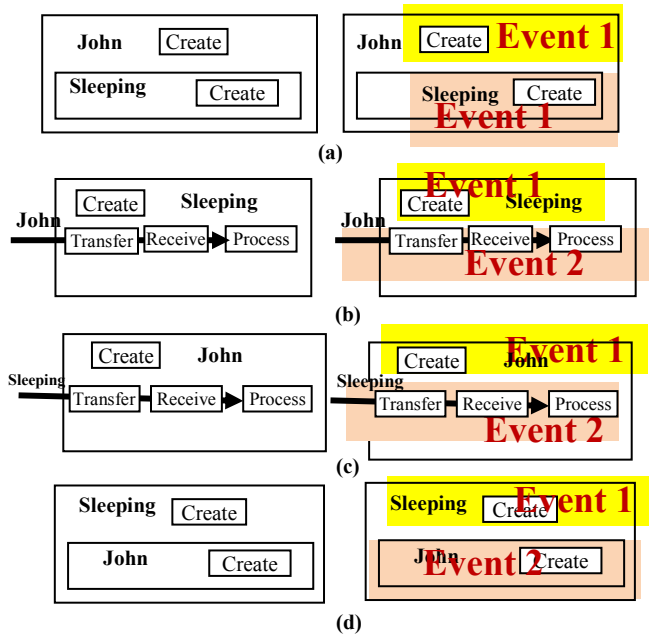

Fig. 15. Different chronologies of events related to John and sleeping.

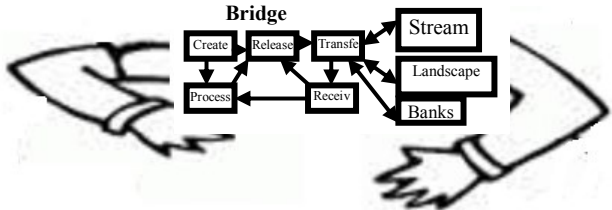

Fig. 16. A thing is a gathering of its constituents.

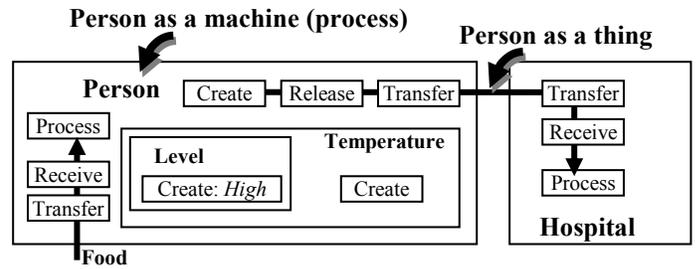

Fig. 17. A person "gathers" as a thing and as a machine.

Notice that, in Fig. 16, a bridge is a thing because it is the object of creating, processing, releasing, transferring, and receiving. It is a process because it is a subject that can create, process, release, transfer, and receive other things. Thus, the TM system consists of changes in processes as the result of flows of those processes. As illustrated in Fig. 17, a person is a "gathering" phenomenon as a thing and as a process. A thing is viewed as an integral object because of gestalt effects when there is no interest in its internal parts (submachines). A person is a thing, but in the eyes of his or her doctor, he or she is packaged with circulatory, respiratory, digestive, excretory, nervous, and endocrine machines. Accordingly, the "reach/boundaries" in Heidegger's things [59] are defined by their thing-ness and process-ness.

We denote all the subdiagrams (power sets) of a TM diagram as a PTM. Let R be a partition (no overlap) subset of the PTM, called regions. Additionally, we define the following mapping from R to durations of T: (r, t) where r in R and t is a duration in T, and each (r, t) is called an event, denoted by $E_i$. The behavior of TM, E, is defined as an ordered, directed graph with the set of vertices {(r, t)} and a set of edges E.

**Example**: Fig. 19 shows a sample event = (reign, duration of time), where the region is *the retrieval of sum* (i.e., release, transfer, transfer, receive).

Fig. 20 shows one of the partitions of the formula diagram broken into events. Fig. 21 shows the behavior of the formula in terms of its events.

### III. TM DEFINITION OF BEHAVIOR

A TM graph is as formal as a formula. Consider the formula shown in Fig. 18: the figure shows its corresponding TM representation. In Fig. 18, the sum and *i* (circles 1 and 2) are added (3) to create the result (4) that is sent to the sum. To monitor the number of additions, *i* is incremented (5) and checked if it has reached *n* (6); if it has, then the result is output (7). The dashed arrow indicates triggering.

Fig. 18 as a static description is free of commitment to behavior and time. It establishes the functions of different submachines in the context of the whole model. On the other hand, a dynamical system in a TM involves time T that is mapped to the TM graph.

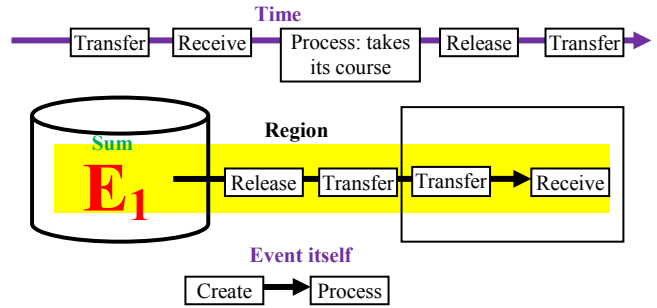

Fig. 19. The event *Sum is retrieved*.

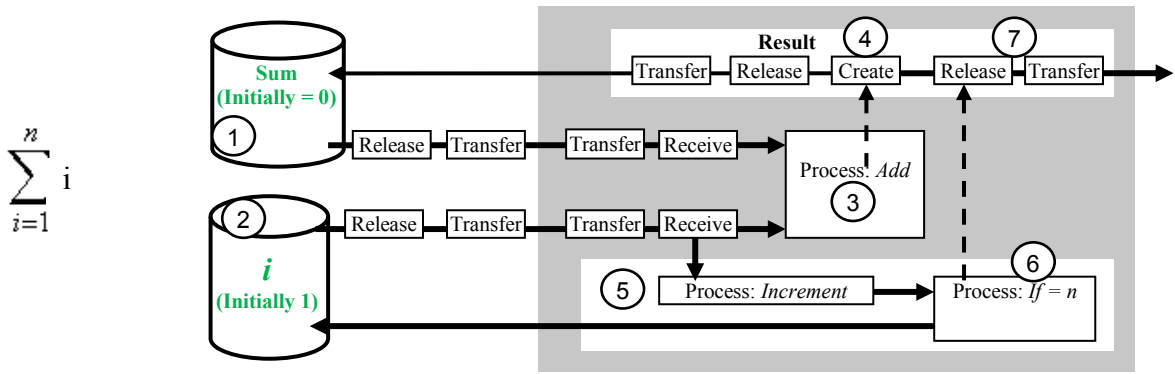

Fig. 18. A sample formula and its TM representation.

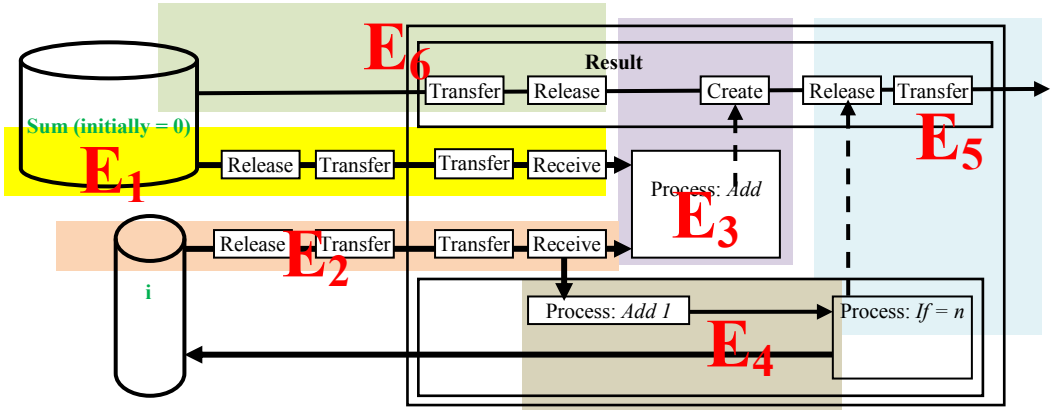

Fig. 20. A partition of regions in the formula example.

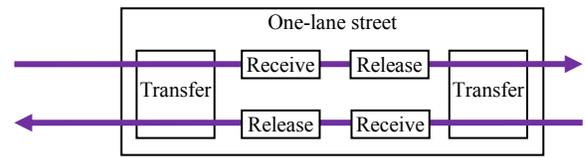

Fig. 22. Model of one-lane street with two opposite directions expressed in a TM.

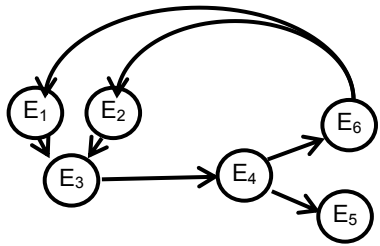

Fig. 21. The behavior of the formula.

## IV. STATIC VS. DYNAMIC TMS

The static model (without events) may include contradictions, as in a one-lane street that permits flows in both directions (Fig. 22). The dynamic description (see Fig. 23) solves that contradiction by creating events based on time (AM and PM). The event indicates the presence of a now (time) and a here (space), as shown in Fig. 23 (top). Fig. 23 (bottom) shows the chronology of events. This diagram is the so-called state diagram.

## V. STATES AND BEHAVIOR: EXAMPLE 1

Bock and Odell [60] illustrate behavior using state diagrams of a factory operation for changing the color of an object, whereby each time the behavior happens is considered to be a separate occurrence, usually at different times, involving different objects and colors (see Fig. 24). According to Bock and Odell [60],

> The occurrences following the three behaviors in [Fig. 24] happen to be the same, but this is only clear from the semantics of the languages, rather than the syntax in [Fig. 24]. For example, the figure . . . [implies] painting must complete before drying starts, even though many explanations of these languages assume it is understood. The semantics is more apparent from the occurrences in [Fig 25].

The solution is shown in Fig. 25, which describes the behaviors on the vertical axis, time on the horizontal, and occurrences as interval bars on the graph.

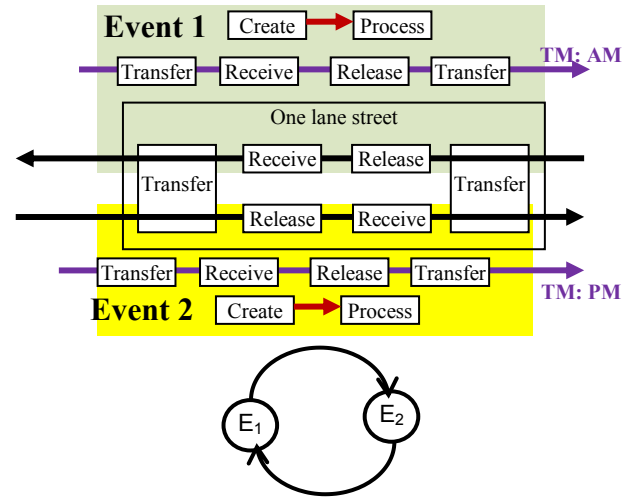

Fig. 23. The behavior in a one-lane street with two opposite directions.

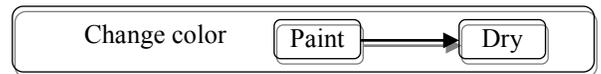

Fig. 24. Paint-drying diagrammatic model (adapted from [60]).

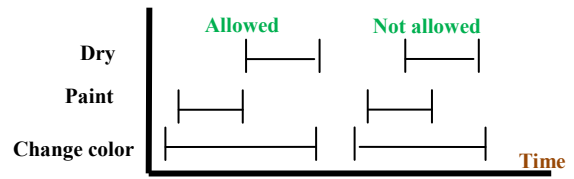

Fig. 25. Multiple behaviors (adapted from [60])

Fig. 26 shows the TM model (static, dynamic, and behavior descriptions) of this example.

We can also control the behavior of the functional machine by including control machines. Suppose that we want to control the quality of painting and drying shown in the example. Fig. 27 shows two additional machines that check the quality of color after painting is completed. If the color is not rated according to a specified level, then the object is sent back to have another coat of color put on; otherwise, the colored object is passed to the drying phase. Similarly, there is another machine to test the dryness level.

Additionally, Bock and Odell [60] give an example of multiple behavior generation where the same occurrence can follow multiple partially specified behaviors, as illustrated in [Fig. 28] for the two behaviors at the top. Fig. 29 shows the corresponding TM model. We assume that the cleanup process concerns the paint brush. Fig. 30 shows four selected events.

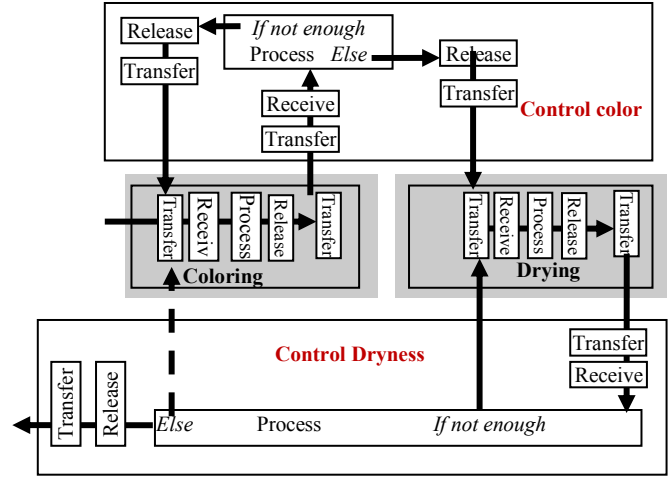

Fig. 27. Control of functional behavior.

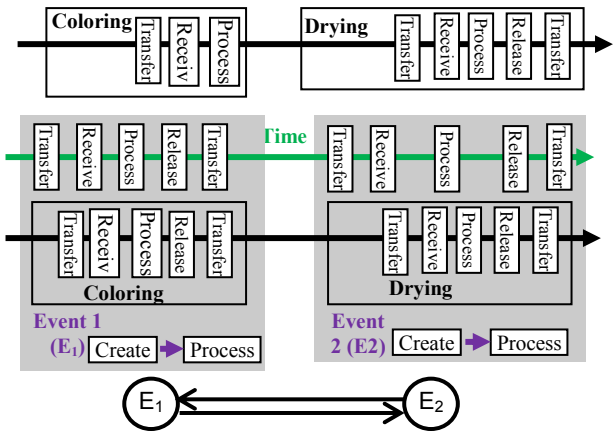

Fig. 26. Paint-drying TM model (Top: static representation; middle: events; and bottom: behavior.

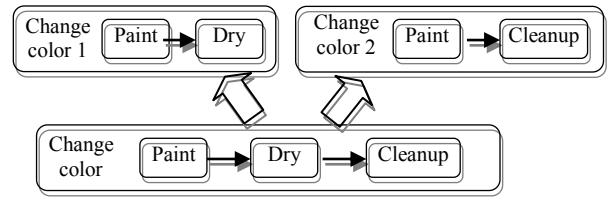

Fig. 28. Multiple behaviors (adapted from [60])

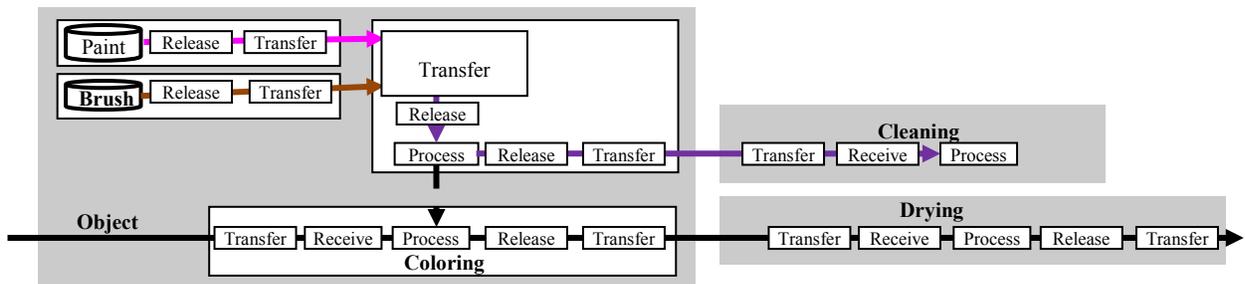

Fig. 29. The TM static description of the multiple behaviors.

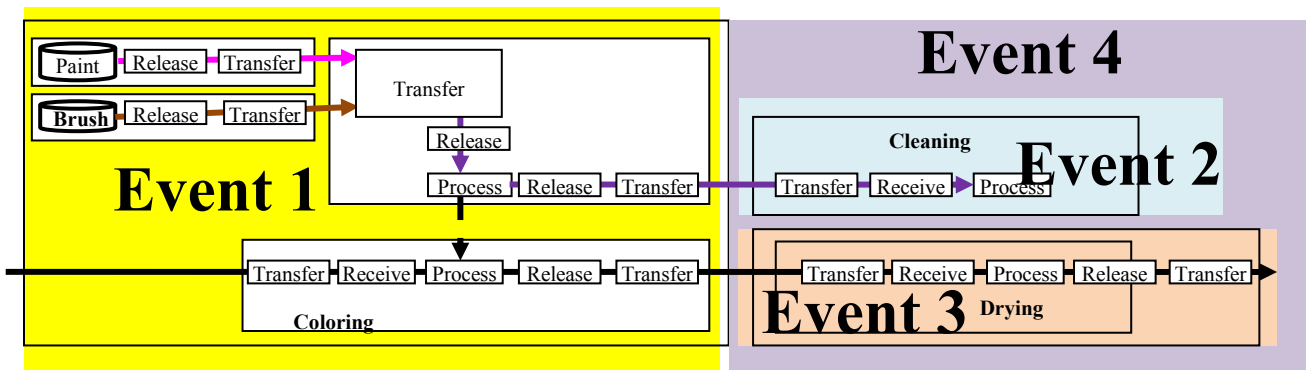

Fig. 30. The events of the multiple behaviors.

Fig. 31 shows the behavior of a system that permits multiple behaviors. Event sequences can be Event 1 → Event 2, Event 1 → Event 3, and Event 1 → Event 4.

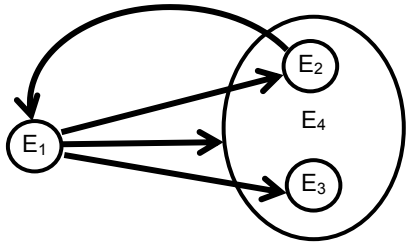

Fig. 31 The chronology of events.

## VI. STATES AND BEHAVIOR: EXAMPLE 2

According to Easterbrook [61], objects have states, and if an object exists, then it has a value. Each possible assignment of values to attributes is a state. An object's nonexistence is also a state. Fig. 32 shows a stacked object's state diagram.

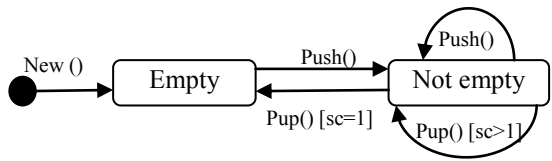

Figure 32. State diagram of a stack (re-drawn, partially from [62]).

### A. Static TM Representation

Fig. 33 shows the TM representation of a stack. We assume that the stack is located in the array Stack [0], Stack [1], Stack [2], ... and that the stack is empty when top = -1. The stack's process starts when the user selects the operation to perform: push or pop (0 in the figure). Accordingly, the diagram can be described as follows.

**Push**: A new item is received (1) to trigger the retrieval of the top value (2), which is incremented (3), and the new top value is stored (4). A record is constructed (5), including the new item (6) and the new top value (7), which is then sent to the storage system (8). There, the record is processed (9), and the new top is used to store the new item (10) in the stack. Note that step (circle) 9 extracts the top and the item from their record constructed in step (circle) 6.

**Pop**: The pop signal triggers (11) the examination of the top value (12). If that value is less than 0 (13), an error message is sent. Otherwise, the following actions occur:

• The top value is decremented (14), and a new value is stored (15).

• The top value is sent to the storage system (16).

In the storage system, the top value is used (17) to retrieve the top item in the stack (18) and to output it (19).

### B. Events

The TM-based modeling of the stack's dynamic aspects provides an alternative way to specify events, as previously defined. To identify the events in the stack example, Fig. 34 shows selected events as follows:

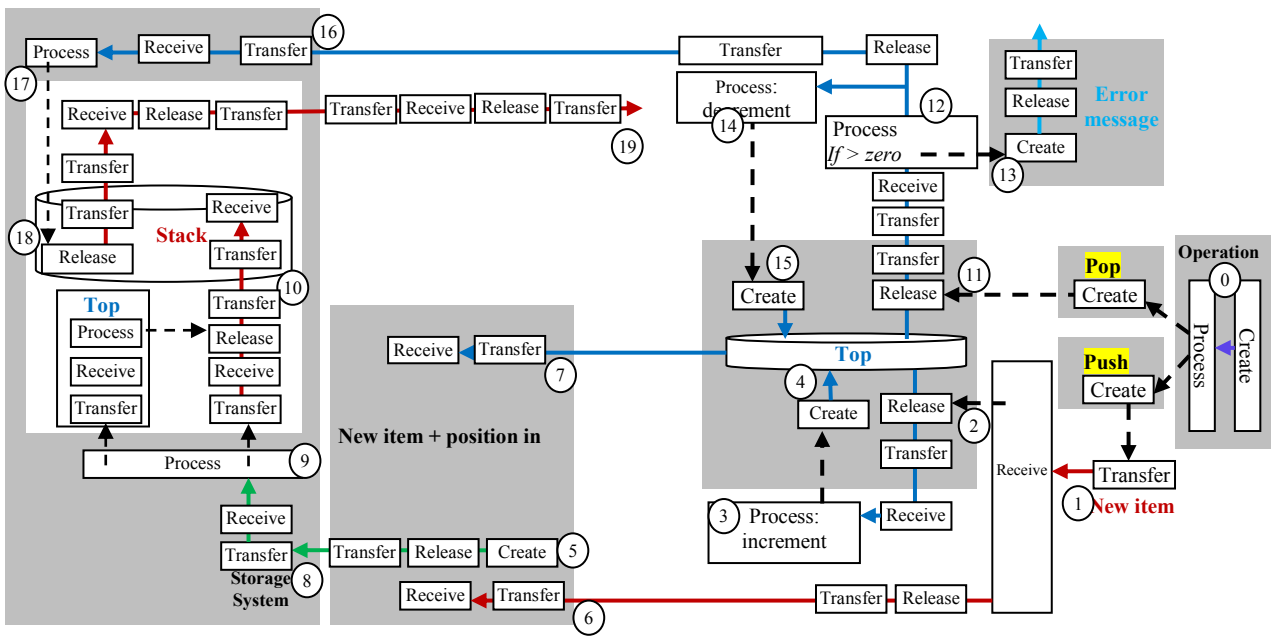

Fig. 33. The flow machine representation of a stack.

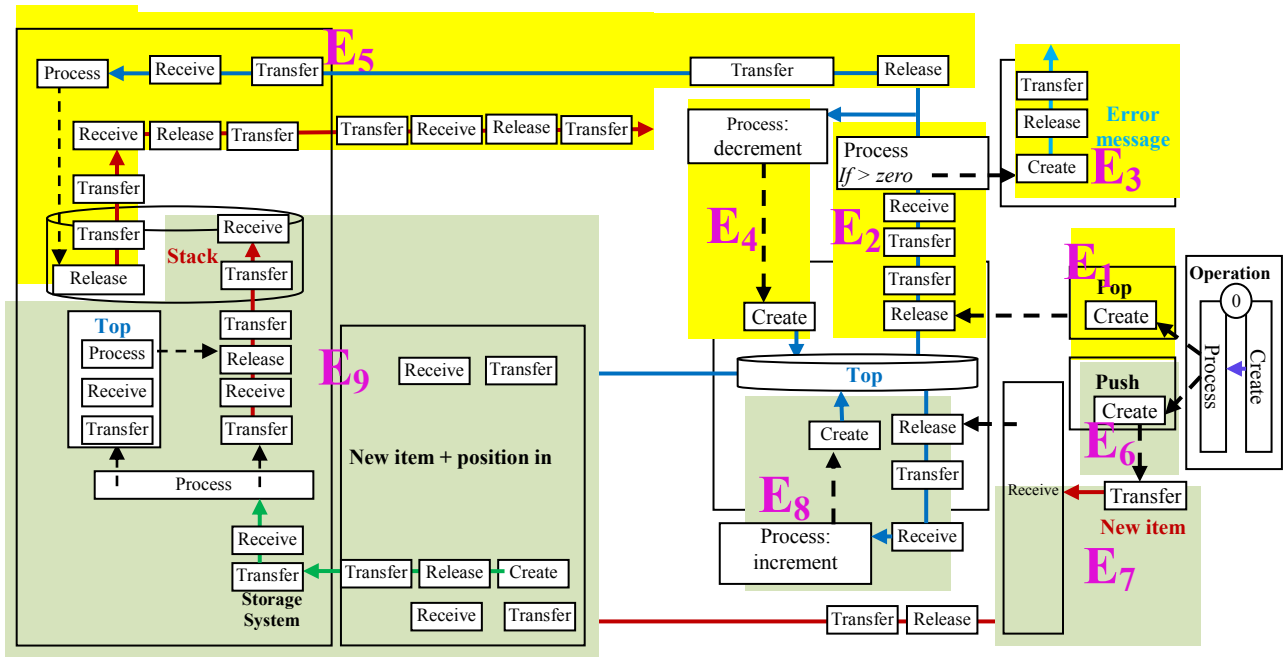

Fig. 34. The events in the TM representation of a stack.

Event 0 ($E_0$): An operation (push or pop) is selected.
Event 1 ($E_1$): A pop is created (generated).
Event 2 ($E_2$): Processing the top value reveals that it is negative, so an error occurs.
Event 3 ($E_3$): Processing the top value reveals that it is positive.
Event 4 ($E_4$): The top value is decremented.
Event 5 ($E_5$): The current top item of the stack is retrieved and sent to the user.
Event 6 ($E_6$): A push is created.
Event 7 ($E_7$): A new item is received.
Event 8 ($E_8$): The top value is incremented.
Event 9 ($E_9$): The new item is put on the stack.

We assume that Event 0 is the event of selecting push or pop.

Therefore, the execution of the stack processes is implemented according to the chronology of events shown in Fig. 35, which reflects the dynamic behavior of a stack in which each event can be considered an execution module (e.g., a programming function or subprogram). Let $E_i$ () denote the module, with its parameters inside the parentheses. In this case, $E_0$ () calls either $E_1$ () or $E_6$ (). If it is $E_1$ () (i.e., in pop), then E1 () calls $E_2$ (). $E_2$ () calls either $E_3$ () or both $E_4$ () and $E_5$ () (i.e., top), depending on the top value. If it calls $E_3$ () (i.e., if top is negative), then there is a printing error, and the program goes back to $E_0$ (). If it calls $E_4$ () and $E_5$ (), then the top value is updated, and the item on the top of the stack is retrieved and sent to the user.

Likewise, if $E_0$ () calls $E_6$ (), then $E_6$ () calls $E_7$ () to increment the top value; it also calls $E_8$ () to receive the new item. Then, $E_8$ () calls $E_9$ () to put a new item on the top of the stack. $E_9$ () would perform that action only after ensuring that the top value has been incremented.

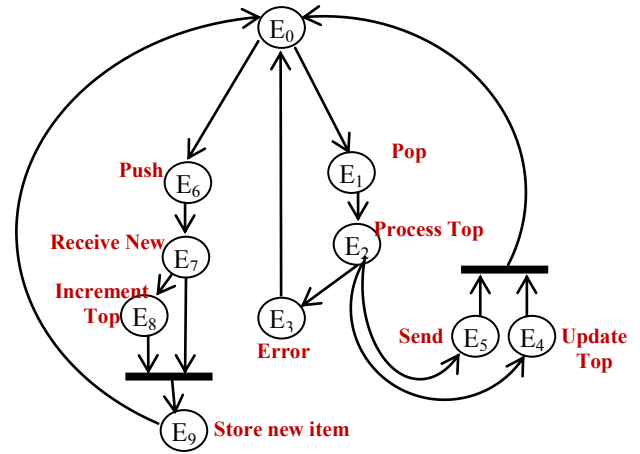

Fig. 35. The execution of events of a stack.

VII. CONCLUSION

In this paper, five generic elementary processes (creating, processing, releasing, receiving, and transferring) are used to form a unifying higher-order process (i.e., a TM) that is utilized as a template in modeling the behavior of systems. A question regarding these five generic processes needs to be addressed: Can all processes be expressed in terms of them? The analysis of behavior in this paper seems to support such a thesis. Additionally, many systems in various fields of study have been applied to TM-based modeling. Further research is needed in this area. Meanwhile, a TM seems to be a valuable tool for analyzing systems, as demonstrated in this paper with regard to the notion of behavior.